\documentclass[aps,prl,superscriptaddress,reprint]{revtex4-1}
\usepackage{graphicx}
\usepackage[permil]{overpic}
\usepackage{latexsym}
\usepackage{amsmath}
\usepackage{amssymb}
\usepackage{upgreek}
\usepackage{xcolor}
\begin{document}
\title{Phenomenological model of supercooled liquid as a possible resolution of the Kauzmann paradox}
\author{M. V. Kondrin}
\email{mkondrin@hppi.troitsk.ru}
\affiliation{Institute for High Pressure Physics RAS, 108840 Troitsk, Moscow, Russia}
\author{Y.B. Lebed}
\affiliation{Institute for Nuclear Research RAS,  Moscow, Russia}
\author{A.A. Pronin}
\affiliation{General Physics Institute RAS, 117942 Moscow, Russia}
\author{V.V. Brazhkin}
\affiliation{Institute for High Pressure Physics RAS, 108840 Troitsk, Moscow, Russia}
\begin{abstract}
The diverging relaxation time in approaching hypothetical ideal glass transition is a subject of hot debate. In the current paper we demonstrate, how diverging relaxation time and turning excess entropy to zero (which is an essence of Kauzmann's paradox) can be avoided, using as an example the model molecular glassformer, propylene carbonate. For this purpose we compare its thermodynamic and dielectric relaxation properties,  both known from the literature. The agreement between two sets of data can be achieved, if we suppose, that enthalpy of supercooled liquid propylene carbonate is governed by activation law, and relaxation time follows double exponential law. We propose the generalized Adam-Gibbs law to reconcile this two dependencies, and qualitatively discuss its implications.
\end{abstract}
\maketitle
\begin{figure}
\includegraphics[width=\columnwidth]{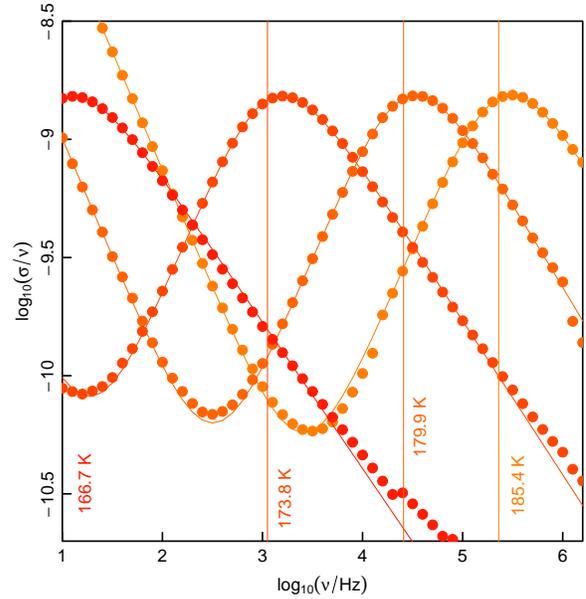}
\caption{Imaginary part of dielectric relaxation (loss function) in propylene carbonate at different temperatures according to Ref.~\cite{kondrin:jcp12}. The symbols are experimental data, solid curves are fits to model relaxation function. Vertical lines mark positions of characteristic frequencies obtained from the fits.}
\label{dielectric} 
\end{figure}

The most fundamental characteristic of liquid (supercooled or normal) is the relaxation time $\tau_0$, which separates phonon-like vibrations at  frequencies higher than $\omega_0 \propto 1/\tau_0$ from slowly relaxing shear movements at lower frequencies. It should be kept in mind, that this border is not sharp, rather $\tau_0$ or $\omega_0$ is f characteristic time/frequency,  describing certain distribution of relaxators (Fig.~\ref{dielectric}). In normal liquid this relaxation is in the range below 100 GHz, but rapidly (superactivationally) shifts into the lower frequency region with lowering the  temperature  ( as the liquid becomes supercooled). So, in the vicinity of glassification temperature (defined by convention, as the temperature, at which relaxation time is below the typical measurement time, that is $10^2-10^3$ s), the  distribution can be conveniently measured in supercooled liquid by dielectric spectroscopy technique (if the liquid is polar, see Fig.~\ref{dielectric}), by shear modulus measurements \cite{schroter:jncs02} or ultra sound attenuation techniques \cite{kondrin:jcp12}. In most liquids these methods  give similar dependence of relaxation time vs. temperature \cite{schroter:jncs02}. A notable exception (for not well known reasons) is monoalcohols \cite{bohmer:pr14,danilov:jpcb17}, but for propylene carbonate, which will be used as an example throughout this paper, the coincidence of relaxation time, measured by different  techniques,  is well established. Additional advantage of propylene carbonate in comparison to other popular molecular glassformers (such as  glycerol or propylene glycol) is, that propylene carbonate retains unimodal relaxation till rather high pressures (4.6 GPa \cite{kondrin:jcp12}), in contrast to glycerol and propylene glycol, whose relaxation at these pressures splits \cite{pronin:pre10,pronin:jetpl10,kondrin:jpcb18}, thus they can not be regarded as simple liquids. Moreover, propylene carbonate has relatively high boiling temperature (513 K) and correspondingly high crossover temperature T$_A \approx 300$ K \cite{stickel:jcp96} (for comparison, glassification temperature of propylene carbonate T$_g$=159 K) at which crossover from high-temperature Arrhenius dependence to low temperature super-activation one occurs. This fact makes  possible testing various analytical models in comparatively wide temperature range,  to describe this superactivation dependence. 

Note, that the slower the liquid measurement experiment is, the lower the temperature of liquid vitrification. This is reflection of the fact, that slower experiments are able to measure the liquids with slower relaxation times, so at extremely long time, the glass will behave  as a liquid itself \cite{brazhkin:jetpl20}. What happens with the relaxation time, when the liquid vitrifies? There is no clear answer, but certainly the liquid (glass) becomes non-ergodic, and the sample more or less freezed in the state, corresponding to the temperature, at which glassification took place. Certainly, some sort of relaxation exists even in the glassy state, which manifests itself in the experiments on glass aging \cite{morvan:jncs21}. These experiments demonstrate, that the properties of glass, kept for a long time at a fixed temperature below the glass transition temperature, are very different from those of fresh glass (obtained in the single experimental run from glassification temperature to this fixed temperature). 

Manifestations of vitrification are manifold. Besides the observation, that the relaxation process shifts below some fixed frequency, it can be observed as a drop by the typical value of several $k_B$ per {\em molecule} in the system of the liquid's specific heat   \cite{sugisaki:bcsj68,angell:jml93}. These processes are not quite different. The drop of specific heat in the glassy state, compared to supercooled liquid, may be associated with the shift of slow movements into nonergodic region, so that subsequent measurements of the specific heat   involve only fast vibrations. It is indeed observed in experiments for high frequency measurements of specific heat near the  temperature of vitrification. It was concluded, that specific heat in the high frequency asymptotic drops to values of about the glassy state values \cite{birge:prl85}. Vibrational properties of glass are difficult to estimate, and even more difficult  to calculate, but it turns out, that  integral characteristic of glass phonon vibrations, such as phonon specific heat, is remarkably close to that of crystal phase \cite{trachenko:prb11}. It can be thought of in this way: phonon density of state of the supercooled liquid/glass  can be regarded as the broadened and slightly shifted to the lower region version of density of state of the crystal, except for the  frequency region close to zero. Since the density of state in the region below 100 GHz (a typical relaxation frequency of supercooled liquid) is low and tends to zero, the difference between vibrational/phonon specific heat of supercooled liquid and crystal is negligible. As a rule of thumb, this vibrational contribution to the specific heat $c_V$ near the melting temperature can be evaluated as being significantly lower than 3$k_B$ per {\em atom} in the system \cite{sugisaki:bcsj68,angell:jml93}. So, the excess (let us name it that) specific heat of slow movements which ``disappear'' during vitrification process in glassy molecular liquids (where molecules consists of several atoms)  makes  only a small contribution into  total specific heat of the glass/liquid. Rough method of evaluation of  the specific heat drop is to subtract the specific heat of the crystal from that of the supercooled liquid in the temperature region, where the measurements of both quantities are possible, and extrapolate it to adjacent temperature regions. Supposedly, this method should yield a good approximation of excess specific heat, although it turns out to have a slight temperature dependence. 

Now we (following Kauzmann) envisage a sort of thought experiment -- what happens with specific heat and internal energy of the supercooled liquid, if we infinitely slow down our cooling process? Certainly (in light of the  above said), glassification temperature will shift to lower temperatures, so we can probe even slower relaxation processes. Kauzmann \cite{kauzmann:cr48,stillinger:jcp88,angell:jrnist97} supposed, that the specific heat drop $\Delta c_P$ during vitrification process at slower cooling rates does not change much (it is almost constant). If we take experimental heat of fusion $\Delta H$ and divide it by this specific heat drop during vitrification process, it turns out, that the difference of internal energy between supercooled liquid and crystal should vanish at some non zero temperature $T_K=T_m-\Delta H/\Delta c_P$  (energetic Kauzmann's temperature), where $T_m$ is the melting temperature. Kauzmann discovered, that for practically all supercooled liquids, this temperature  is above zero. From a simple consideration it also follows, that the specific heat in this point should change discontinuously, so this transition (if achieved) was believed to be similar to the second order.

Besides energetic Kauzmann's temperature, there exists an entropic one (where excess configurational entropy drops to zero), which causes more problems, since it is closer to the melting temperature, than the energetic Kauzmann's temperature. It can be understood from the simple inequality: 
 
\begin{equation}
\frac{\Delta H(T_m)}{T_m} = \Delta S = \int\limits_0^{T_m} \frac{c_PdT}{T}>\int\limits_0^{T_m} \frac{c_PdT}{T_m}=\frac{\Delta H(T_m)-\Delta H(0)}{T_m}
\end{equation}

So, the entropic Kauzmann's temperature is the one, at which only some part of enthalpy of fusion turns to zero (in contrast to energetic Kauzmann one where all enthalpy of fusion drops to zero). This inequality also demonstrates, that in supercooled liquid freezed in infinitely slow rate even at zero energy, a residual internal energy $\Delta H(0)$ should exist. This is a simple reflection of the fact, that supercooled liquid at zero temperature is in metastable state, compared to  crystal.

Adam and Gibbs assumed, that reaching the Kauzmann temperature requires infinite time, and proposed the formula between relaxation time at fixed temperature $\tau(T)$  and excess entropy of the liquid $\Delta S(T)$ at the same temperature $\tau(T)=\exp\left(\frac{B}{T\Delta S(T)}\right)$ \cite{adam-gibbs:jcp65}. By the order of magnitude $T\Delta S(T)$ is equal to enthalpy of the supercooled liquid $T\Delta S(T) \approx \Delta H(T)-\Delta H(0)$ (it can be shown later, that in our model this equation holds true with good accuracy). At Kauzmann's temperature $\Delta H(T)-\Delta H(0)$ is zero, so the relaxation time at finite temperature should be infinite. It was supposed, that this notion corresponds to experimental data, because previously relaxation time observed in supercooled liquids was fitted by the Vogel-Fulcher-Tamman (VFT) equation $\tau(T)=\exp\left(\frac{C}{T-T_0}\right)$ where $T_0$ was believed to be close to Kauzmann's temperature. Although this   approximation is good in the case of propylene carbonate \cite{angell:jml93},  for many other glassformers more recent studies provide little evidence of relaxation time diverging  at finite temperatures \cite{hecksher:np08}. Clearly, there is a problem  of how to reconcile the non-diverging relaxation time at finite temperature  with non-zero thermodynamic Kauzmann's temperature (which can be concluded from experiments), at which  this time is supposed to diverge.

\begin{figure}
\includegraphics[width=\columnwidth]{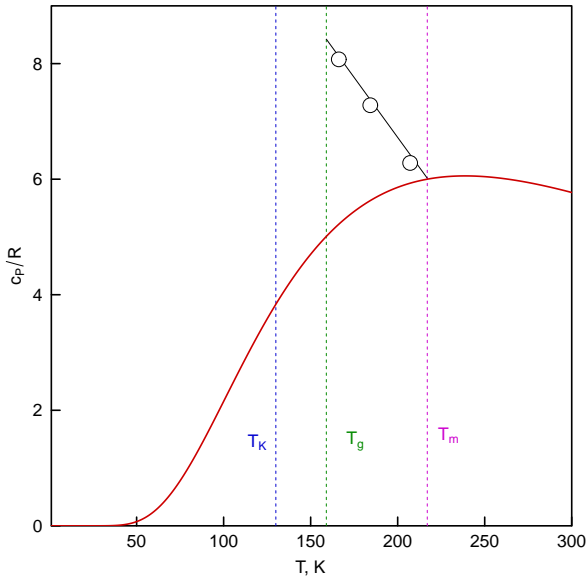}
\caption{The difference of specific heat between supercooled liquid and crystal states of propylene carbonate. Red curve is a theoretical approximation according to activation law, $\circ$ -- experimental data of specific heat difference  (Angell {\em et al.} \cite{angell:jml93}). $T_m$, T$_g$ and $T_K$  are experimental melting, glassification and Kauzmann's temperatures. }
\label{cp}
\end{figure}

The simple idea is, that  excess specific heat and excess internal energy of supercooled liquid should  tend to zero smoothly at low temperature. Let us consider the case, that excess internal energy of liquid is governed by activation law : $\Delta H (T) - \Delta H (0) = H_0 \exp\left(-\frac{E}{T}\right)$ (the energy and temperature we will treat in energy units $k_B=1$). Then the specific heat of this contribution to liquid energy can be obtained as : $\Delta c_P=\frac{H_0 E}{T^2}\exp\left(-\frac{E}{T}\right)$. As a side remark, one can recall the Ref.~\cite{trachenko:prb11} where the drop of specific heat at constant pressure during glassification processes related to the ratio between thermal expansion coefficient $\alpha$ and bulk modulus $B$ in the supercooled liquid and glassy state in the well-known thermodynamic equation $c_p-c_v=-\alpha^2TBV$ which can reach values of about 100 \% of specific heat in glass state.

To resolve this set of equations using experimental data for 3 unknown parameters ($\Delta H (0)$, $H_0$ and $E$), we need the third equation which is provided by the balance of entropy:

\begin{equation}
\Delta S(T_m)=H_0\int\limits_0^{T_m}\frac{d \exp\left(-\frac{E}{T}\right)}{T}=H_0 \exp\left(-\frac{E}{T_m}\right)\left(\frac{1}{T_m}+\frac{1}{E}\right)
\end{equation}

For propylene carbonate we can solve equations for specific heat and entropy taking the experimental values of the excess specific heat at melting temperature ($\Delta c_p \approx 6R$) and entropy of fusion ($\Delta S(T_m)=\Delta H(T_m)/T_m \approx 4R$. They can be easily resolved and yield the values $E= T_m\frac{\Delta c_p}{2 \Delta S(T_m)}\left(1+\sqrt{1+\frac{4 \Delta S(T_m)}{\Delta c_p}}\right) \approx 470 $ K.  From the equation for enthalpy, other parameters can be evaluated (for resulting curve see Fig.~\ref{cp}), in particular, residual enthalpy at zero temperature is approximately equal to $1/3\Delta H(T_m)$. This temperature dependence predicts the increase of $\Delta c_P$ with temperature lowering above melting temperature, but, on the other hand, significantly underestimates the experimentally observed dependence (calculated as the difference between the specific heat of the supercooled liquid and the crystal, see Fig.~\ref{cp}). This can be due either to the fact, that the true temperature dependence is stronger than the simple activation dependence, or to approximate character of the specific heat difference between the liquid and crystal (more precise approximation would be the difference between liquid and glass, although we acknowledge significant  experimental  difficulties of obtaining these values). 

\begin{figure}
\includegraphics[width=\columnwidth]{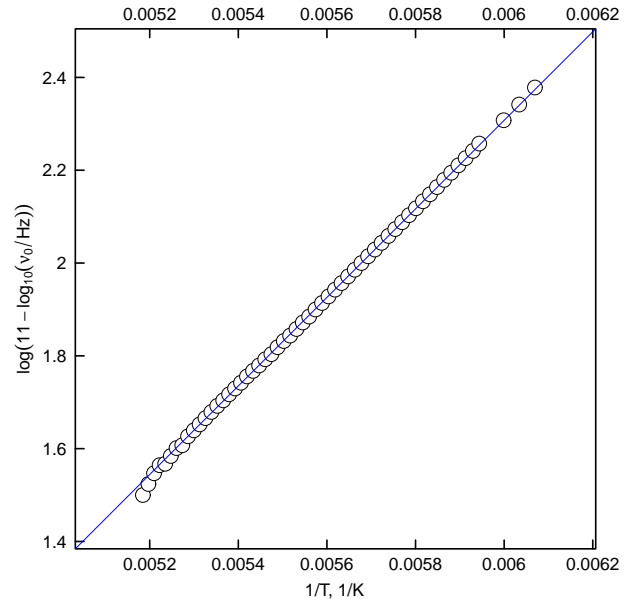}
\caption{Dielectric relaxation time in propylene carbonate vs. temperature \cite{kondrin:jcp12}. Blue line is a fit according to the Bradbury-Shishkin equation  (Eq.~\ref{bs}).}
\label{relax}
\end{figure}

\begin{figure}
\includegraphics[width=\columnwidth]{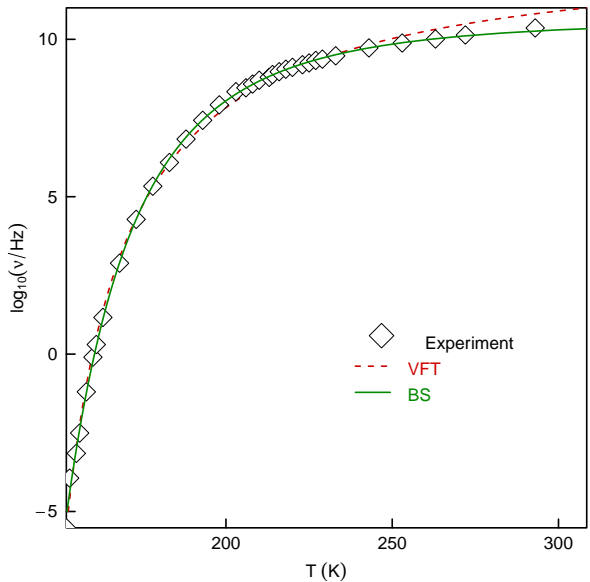}
\caption{Dielectric relaxation time in propylene carbonate vs. temperature \cite{schneider:pre98,lunkenheimer:cp00,lunkenheimer:prl05,lunkenheimer-kastner:pre10} (symbols). Dashed red line is the fit of experimental data by VFT equation, solid green line -- by the Bradbury-Shishkin one (Eq.~\ref{bs}).}
\label{relax1}
\end{figure} 

How will behave the relaxation time in the same temperature region? For activation law the  characteristic frequency of dielectric relaxation follows the double exponential  (the Bradbury-Shishkin ) law \cite{bradbury:tasme51,shishkin:ztf55,sanditov:82,sanditov:gpc02,sanditov:pss16,sanditov:pu19})(see Fig.~\ref{relax}): 
\begin{equation}
\nu=\nu_0 \exp(-A\exp(E/T))
\label{bs}
\end{equation}
This equation seems to be reinvented many times since then (see e.g. Ref.~\cite{angell:jpc72}), and describes non-diverging relaxation time at finite temperatures . It can be regarded as slight modification of the  Waterton-Mauro equation \cite{waterton:jsgt32,mauro:pnas09,lunkenheimer-kastner:pre10} (also known as MYEGA). It results from application of Adam-Gibbs relation to the internal energy, following the activation law. We should note, that in  limited temperature range (as depicted in Fig.~\ref{relax}), it is not evident that the Bradbury-Shishkin equation   is preferrable than its counterpart (VFT). In fact, it is slightly worse, and its determination coefficient $R^2$ is marginally lower, than that of VFT equation (0.9996 {\em vs.} 0.9998 respectively). The situation drastically changes, if we fit the data obtained in the significantly wider frequency range,  including THz measurements, as well as data from aging experiments \cite{schneider:pre98,lunkenheimer:cp00,lunkenheimer:prl05,lunkenheimer-kastner:pre10}. Although in this data, there was slightly different convention of characteristic frequency adopted  (inverse ``average'' relaxation time), in contrast to those used by us before (from the fit of relaxation response), we still consider the difference  insignificant. Fit of this experimental data by VFT and BS equation is shown in Fig.~\ref{relax1}. Without thorough statistical analysis it is evident, that BS equation provides better fit of experimental data. It corresponds to the conclusions of Ref.~\cite{lunkenheimer-kastner:pre10}, that MYEGA equation provides better fit of broad band relaxation data, obtained on propylene carbonate, than VFT one. So, in our fit we just corroborate observation of Ref.~\cite{sanditov:pu19} that all ``double exponential'' equations provide approximately equally good fits of experimental data.

However, there is a problem, that the activation energy $E=960$ K, obtained from the dielectric relaxation time is almost two times higher, than the one deduced from thermodynamic consideration (470 K). This discrepancy can be reconciled, if we take into account qualitative character of the Adam-Gibbs relation. Instead, one should propose a generalized Adam-Gibbs relation where the  relaxation time is related to  internal energy of the liquid $H(T)=\Delta H (T) - \Delta H (0)$ by the equation:

\begin{equation}
\tau \propto \exp\left(\left(\frac{B}{H(T)}\right)^\gamma\right)
\label{gag}
\end{equation} 

It is clear, that in propylene carbonate the exponent $\gamma$ should be close to 2. 

There should be short interpretation of the generalized Adam-Gibbs equation (\ref{gag}). Activation law $\Delta H(T)- \Delta H(0)$ is suitable for description of concentration of some walls,  separating different molecular flows. These walls require  a certain energy $E$ for their creation. Once created, they separate the whole volume into ``rigid blocks'', with flows along these walls as the single entities.  Since the flows are constricted by the tubes of current, the exponent $\gamma \approx 2$. However, the flow of these entities is hampered inversely exponential to their size, according to the generalized Adam-Gibbs law (\ref{gag}). Although this model seems to lack direct evidence, it is not inferior to other models of relaxation processes in supercooled liquids.  

To conclude, we provide phenomenological foundation of non-divergence of  relaxation times in propylene carbonate. Comparison of its' thermodynamic and kinetic properties  (relaxation frequency obtained through broadband dielectric spectroscopy) can be satisfactorily described, using notion of activation character of internal energy resulting in the ``double exponent'' relaxation time behavior, which can be described by the generalized Adam-Gibbs relation Eq.~(\ref{gag}). This allows us get rid of Kauzmann's paradox.

This work is supported by the Russian Science Foundation (grant No. 19-12-00111).

%

\end{document}